# Knowledge and Influence of MOOC Courses on Initial Teacher Training

**Jessica Pérez-Parras** [a] **and José Gómez-Galán** [b]

**Abstract**: The impact of MOOC courses in the processes of distance learning has been extremely important from the very beginning. They offer an innovative model of massive teaching, which exploits in a paradigmatic manner the potential and relevance that ICT's currently have in modern society. The present article has as its primary objective the analysis of the presence of these courses and the role that they represent in teacher training, and their knowledge and influence on the future teachers that are currently being formed at university level. A case study has been carried out with descriptive not experimental methodology, from a quantitative base. The sample study has been undertaken in Spain (n=200). Its main result being the determination of the minimal impact that the MOOC phenomenon has had on the students polled. Equally, a significant lack of knowledge has been revealed in all its dimensions (professional, pedagogical, structural, etc.), with only a minority of those in the sample group having indicated that they have studied any of the courses, or know to some extent the main platforms of the world in which they are offered. A large number of those surveyed therefore are unaware of the existence of these courses. As a result, it has been established that, regardless of the quality of the learning and the didactic and methodological characteristics that the MOOC courses offer, their study and analysis is considered necessary for future educational professionals. It is imperative that at the level of Higher Education, and especially in the faculties of teacher training, that the most recent advances in the field of ICT's are introduced in the study plan and in the academic programs, for they constitute the base of modern society. This will permit not only the granting of technical abilities to university students, future professionals, but will also permit the undertaking of a critical analysis of the characteristics of our world which will contribute to the boosting of everything that is really beneficial for social development.

---

[a] Universidad Autónoma de Madrid (Spain); [b] Universidad Metropolitana (Puerto Rico, United States). Correspondence: Jessica Pérez-Parras, Facultad de Formación de Profesorado y Educación, C/ Francisco Tomás y Valiente, 3, Universidad Autónoma de Madrid, Ciudad Universitaria de Cantoblanco, 28049 Madrid (Spain). jessicaperez1203@gmail.com.

### 1. Introduction

Undoubtedly, Information and Communications Technologies (ICT) today imply profound social transformation, one which needs to be engaged by education. One of the main concerns of the education system is the integration of technologies and media resources into the teaching-learning process as well as the need for a critical analysis to be undertaken into the importance of these resources in our world (Buckingham, 2000 and 2013; McFarlane & Sakellariou, 2002; Gómez-Galán, 2003, 2007 and 2015; Hepp, Hinostroza, Laval & Rehbein, 2004; Schibeci, MacCallum, Cumming-Potvin, Durrant, Kissane & Miller, 2008; Ertmer, Ottenbreit-Leftwich, Sadik, Sendurur & Sendurur, 2012). In this framework, ICT teacher training is fundamental for excellence ICT pedagogical integration (Williams, Coles, Wilson, Richardson & Tuson, 2000; Alonso & Gallego, 2000; Llorente, 2008; Tondeur, Van Braak, Sang, Voogt, Fisser & Ottenbreit-Leftwich, 2012).

Alongside these two important issues there is the continuing emergence of new technological proposals which are emerging in the context of the digital paradigm, and are presenting themselves as not only being innovative but also promise to essentially change the meaning of education in the fields in which they are applied. They are having a significant impact and a great amount of experimental practical and scientific studies are being realized especially in relation to the advantages and disadvantages of their potential, which is essential for the definition and assessment of their true significance (Murphy & Greenwood, 1998; Wood, Underwood & Avis, 1999; Gómez-Galán, 2001 and 2011; Stensaker, Maassen, Borgan, Oftebro & Karseth, 2007; Bingimlas, 2009; Livingstone, 2012; Kreijns, Van Acker, Vermeulen & Van Buuren, 2013; Vrasidas, 2015).

A clear example of this educational-technological impact can be found in MOOC (Massive Open Online Courses) courses. These courses are based on the principles of massive, free access to all materials and resources offered online. This phenomenon has had a major worldwide expansion, opening opportunities at the same time for education and training. In addition to being the entry point for the popularization of science (Anderson y Dron, 2011; Regalado, 2012; Vázquez-Cano, 2013), the future possibilities are enormous and are being studied in all their various dimensions (Adamopoulos, 2013; Aguaded, 2013; Emanuel, 2013; Grover, Franz, Schneider & Pea, 2013; López-Meneses *et al*, 2013; Siemens, 2013; López Meneses, Vázquez-Cano & Gómez-Galán, 2014; Al-Atabi & Deboer, 2014, Gómez-Galán, 2014; King, Robinson & Vickers, 2014; Chen & Chen, 2015; Malchow, Bauer & Meinel, 2015; Reich, 2015; Raffaghelli, Cucchiara & Persico, 2015; Selwyn, Bulfin & Pangrazio, 2015).

The realization of this study is essential because the MOOC paradigm, from its beginnings as an innovative proposal offered by Cormirer and Alexander, has provoked a great revolution in different parts of the world, and produced enormous change and progress in society (Vance, 2013). Many initiatives as a result have been developed to implement this new form of education. The success of some of them, such as Coursera, EdX, Udacity, MiriadaX, among others, has seen Spain and other countries express interest in the implementation of these courses, with several universities already participating in this initiative.

The courses are open, participatory and distributed along a pathway for connection and collaboration as well as job sharing. Some experts consider these courses to be positive while others see them as a threat to current educational systems (Liyanagunawardena, Adams y Williams, 2013; Haynie, 2014; Morrison, 2014). The phenomenon is nevertheless expanding rapidly worldwide, with such speed that the word tsunami is often quoted by some authors when referring to MOOC (Brooks, 2012; Sancho-Salido, 2013), It is therefore imperative to understand how it works internally as well as studying strengths and advantages and the undoubted potential it has for the enrichment of teaching and learning in the 21$^{st}$ century.

## 2. Objectives and Hypotheses of the Study

Once the general approach to the issue has been exposed, the fundamental aim of this research, which focuses on the impact on and the knowledge required by future teaching professionals in Spain with regard to the MOOC phenomenon, will be explained.

From this general objective, several specific goals can be set that will allow an in depth analysis of the subject and the acquisition of additional information. They are: (1) assessing the level of participation that potential teachers have in courses different to MOOC; (2) research into the time elapsed since this phenomenon was first recognised if ever; (3) examining how this phenomenon has come to light (3) analysing the problems found by subjects when running MOOC courses; (4) investigating the choice future professionals must make in relation to languages when signing up for MOOC teaching courses; (5) investigating which of the MOOC platforms are the most used; (6) showing the extent of involvement with other students or professionals throughout the courses; (7) assessing the participation of subjects in these courses; and (8) investigating the motivation that leads respondents to participate in MOOC courses.

Based on the objectives that have been mentioned in the previous section, this research poses a main hypothesis from which several additional hypotheses arise. The main hypothesis is that: future educational professionals in Spain have insufficient knowledge of the MOOC phenomenon and as a consequence the impact upon them has been minimal. On the basis of this main hypothesis other secondary hypotheses have been established : (1) the younger population follows different online courses than MOOC (as they make greater use of technology); (2) the highest percentage of people who are aware of the MOOC phenomenon comprises students who study exclusively (more time employed) the percentage is lower if part or full time work is added to the study pattern (3) MiríadaX is the most widely used platform in Spain (it was created in this country and primarily uses the Spanish language) for the conduction of MOOC courses; and (4) the main reasons leading to the completion of these courses is gratuity (the main feature of these courses).

The research developed into this work will allow us to test the proposed hypotheses, so that information can be obtained pertaining to the necessity for future educational professionals to at least have greater knowledge of MOOC courses or, a thorough knowledge of the phenomenon.

### 3. Methodology and Scientific Study Procedures

This research has been developed employing a non-experimental descriptive methodology. The procedure consists of a literature review which provides information on previous studies and compares that to the results obtained in our study. This also allows information to be obtained on the status of new technologies in today's society as well as the main features of e-learning processes and the initial and ongoing training of teachers in relation to new technologies, so that the emergence of the MOOC phenomenon and more specifically the situation of the phenomenon in Spain can be analyzed.

To perform the empirical framework the collection of information was implemented through a questionnaire involving students in Education degrees at Spanish universities. The sample total was 200 students, of whom 113 were women and 87 men, aged mostly between 18 and 25 years with a smaller percentage between 25 and 35 years. We believe that the sample has been significant enough to obtain objective outcomes.

The main instrument for conducting this research has been the questionnaire by Christensen, Steinmetz, Alcorn, Bennett, Woods & Emanuel (2013), although it has obviously been adapted to the characteristics of education in Spain. Also it was validated through expert opinion before application. It consists of 13 items divided into two parts. The first part is based on general knowledge of the data of the subject under study, while the second focuses on a deeper understanding of the MOOC phenomenon. Each of these items contains various choices, one of which must be selected.

Once this information had been gathered, the statistical analysis of the different sections was initiated with the most appropriate tests being carried out for the goals pursued. Later the preparation of the section on findings and conclusions of the study was carried out. This research not only aims at making contributions to the body of knowledge in this field but focuses on publicizing the MOOC phenomenon, which has had a major expansion globally, as well as within Spain and among future education professionals.

### 4. Data Analysis and Results

To best synthesize this section, each of the questions from the questionnaire used for the study will be detailed as well as the results obtained from each. As we mentioned the questionnaire is divided into two parts, the first of which is based on general data of respondents, whereas the second goes deeper into an understanding of how MOOC works.

#### 4.1. General knowledge of the subject

*General knowledge of the subject*

| | | Frequency | Percentage | Valid percentage | Cumulative percentage |
|---|---|---|---|---|---|
| **Sex** | Men | 87 | 43,5 | 43,5 | 43,5 |
| | Women | 113 | 56,5 | 56,5 | 100,0 |
| | Total | 200 | 100,0 | 100,0 | |
| **Age** | Between 18y24 | 163 | 81,5 | 81,5 | 81,5 |
| | Between 24y35 | 37 | 18,5 | 18,5 | |
| | Total | 200 | 100,0 | 100,0 | 100,0 |
| **Occupation** | Student | 173 | 86,5 | 86,5 | 86,5 |
| | Part-time | 27 | 13,5 | 13,5 | 100,0 |
| | Total | 200 | 100,0 | 100,0 | |
| **Online courses** | Yes | 65 | 32,5 | 32,5 | 32,5 |
| | No | 135 | 67,5 | 67,5 | 100,0 |
| | Total | 200 | 100,0 | 100,0 | |
| **Number of online courses** | Under 5 | 163 | 81,5 | 81,5 | 81,5 |
| | Between 5y10 | 33 | 16,5 | 16,5 | 98,0 |
| | Between10y20 | 4 | 2,0 | 2,0 | 100,0 |
| | Total | 200 | 100,0 | 100,0 | |

Table 1. *General knowledge of the subject.* Source*: Compiled by authors from SPSS Statistical Programme.*

As we can see in *Table 1*, in the selected sample for the study there is a higher percentage of women (56.5%) than men (43.5%). To check the age of the subjects four sample categories have been established: between 18 and 24, between 24 and 35, between 35 and 50 and over 50. The results obtained show that there is a higher percentage of subjects aged between 18 and 24 years (81.5%), and a lower percentage of subjects between 24 and 35 (18.5%) and a nil percentage among the last age groups.

In relation to the current situation of subjects in this sample it is noticeable that most of the subjects (173) are students corresponding to 86.5% of the total, while 13.5% (27) of the remainder, study and work part time simultaneously. With regard to the question as to whether the subjects participated in an online course different to MOOC, it can be seen that 135 subjects (67.5% of the total) that made up the sample group have undertaken no course while 65 subjects or 32.5% have. In relation to the number of online courses, different to MOOC, in which the students have enrolled it is clear that most of the subjects, 81.5% (163 subjects), have studied less than 5 courses, while 16.5% (33 subjects) have enrolled in between 5 and 10 courses. Only 2% (4 subjects) of the sample group studied between 10 and 20.

To test if there is a correlation between the variables of age and participation in online courses different from MOOC, contingency coefficient tests have been completed with no significant results evident. From the figures mentioned above, it can be specified that of the total of 65 people who had attended several courses different to MOOC, 48 are aged between 18 and 24, while the remaining 17 are between 24 and 35 years old. Through Pearson´s *Chi-square* test it has been shown that these differences are not significant.

4.2. MOOC knowledge

| | | *MOOC (A) knowledge* | | | |
|---|---|---|---|---|---|
| | | Frequency | Percentage | Valid percentage | Cumulative percentage |
| **Knowledge of MOOC phenomenon** | Over one year | 10 | 5,0 | 5,0 | 5,0 |
| | Over the past year | 21 | 10,5 | 10,5 | 15,5 |
| | Six months ago | 13 | 6,5 | 6,5 | 22,0 |
| | One month ago | 3 | 1,5 | 1,5 | 23,5 |
| | MOOC phenomenon unknown | 153 | 76,5 | 76,5 | 100,0 |
| | *Total* | 200 | 100,0 | 100,0 | |
| **MOOC knowledge procedure** | Acquaintances | 18 | 6,5 | 27,7 | 27,7 |
| | Internet and social networks | 12 | 5,0 | 21,3 | 48,9 |
| | Direct Advertising | 8 | 3,5 | 14,9 | 80,9 |
| | Others | 9 | 4,5 | 19,1 | 100,0 |
| | *Total* | 47 | 23,5 | 100,0 | |
| Missing values | *System* | 153 | 76,5 | | |
| | *Total* | 200 | 100,0 | | |
| **Problems** | Interaction with instructors | 3 | 1,5 | 11,1 | 11,1 |
| | Video, audio quality | 1 | ,5 | 3,7 | 14,8 |
| | Website browsing | 1 | ,5 | 3,7 | 18,5 |
| | Ineffective forums | 3 | 1,5 | 11,1 | 29,6 |
| | Lack of tiem | 18 | 9,0 | 3,7 | 100,0 |
| | Poor solution and technical assistance | 1 | ,5 | | |
| | *Total* | 27 | 13,5 | 100,0 | |
| Missing values | System | 173 | 86,5 | | |
| | Total | 200 | 100,0 | | |
| **Language** | English | 2 | 1,0 | 7,4 | 96,3 |
| | Spanish | 24 | 12,0 | 88,9 | 100,0 |
| | Others | 1 | ,5 | 3,7 | |
| | *Total* | 27 | 13,5 | 100,0 | |
| Missing values | *System* | 173 | 86,5 | | |
| | *Total* | 200 | 100,0 | | |

Table 2. *MOOC knowledge: problems associated with following these courses and with the selected language.* Source*: Compiled by author from SPSS Statistical Programme.*

In this section an effort was made to establish how many of the subjects surveyed were familiar with the MOOC phenomenon and, if so, for how long. The results suggest that all the subjects in the sample, 76.5% (153 subjects), do not understand the MOOC phenomenon, whereas the remaining 23.5% does. Within this percentage 5% (10 subjects) have had knowledge of MOOC for more than one year, 10.5% (21 subjects) have become familiar with it in the last year, 6.5% (13 subjects) for a period of six months and only 1.5% (3 subjects) have learned about it in the last month.

With regard to the question of how knowledge of the MOOC phenomenon was acquired only 47 were able to respond as the remaining 153 are unfamiliar with it and are considered to be lost cases, as shown below. From the extracted data it can be concluded that 6.4%, i.e., 18 of 47 subjects with knowledge of MOOC, acquired the said knowledge from someone else; 12 (5%) became informed via Internet and social networks; 8 (3.5%) through direct publicity and 9 (4.5%) through other means, specifically teachers.

Of the people who are familiar with MOOC only 27 have made any progress. Three of these 27 people have encountered problems in their interaction with instructors, one with the quality of video or audio, one had problems navigating the website, three considered the Forum was ineffective, 18 felt they had insufficient time to complete their work, and one considers that the solutions are poor and the technical support is inadequate.

In response to the question regarding which language was chosen by subjects when taking these courses, 24 (12%) completed their studies in Spanish**,** and only two in English (1%). One subject referred to "another language" (0.5%) but failed to specify which one.

| | | *MOOC knowledge (B)* | | | |
|---|---|---|---|---|---|
| | | Frequency | Percentage | Valid percentage | Cumulative percentage |
| **Platforms** | Coursera | 7 | 3,5 | 25,9 | 25,9 |
| | EdX | 1 | ,5 | 3,7 | 29,6 |
| | Udacity | 3 | 1,5 | 11,1 | 40,7 |
| | Uned COMA | 2 | 1,0 | 7,4 | 48,1 |
| | Several University Platforms Moodle | 13 | 6,5 | 48,1 | 96,3 |
| | MiríadaX | 1 | ,5 | 3,7 | 100,0 |
| | *Total* | 27 | 13,5 | 100,0 | |
| Missing values | *System* | 173 | 86,5 | | |
| | Total | 200 | 100,0 | | |
| **Interaction** | Yes | 7 | 3,5 | 29,2 | 29,2 |
| | No | 14 | 7,5 | 58,3 | 87,5 |
| | No interaction, but conversation-reading | 6 | 3,5 | 12,5 | 100,0 |
|   | *Total* | 27 | 13,5 | 100,0 |   |
| Missing values | *System* | 173 | 86,5 |   |   |
|   | Total | 200 | 100,0 |   |   |
| **Participation** | Participated and completed | 13 | 6,5 | 27,7 | 27,7 |
|   | Participated and not completed | 14 | 7,0 | 29,8 | 57,4 |
|   | Not completed but I´m interested | 13 | 6,5 | 27,7 | 85,1 |
|   | Not participated and no interest | 7 | 3,5 | 14,9 | 100,0 |
|   | *Total* | 47 | 23,5 | 100,0 |   |
| Missing values | *System* | 153 | 76,5 |   |   |
|   | Total | 200 | 100,0 |   |   |

Table 3. *Platforms used in MOOC. Interaction and participation of subjects.* Source*: Compiled by author from SPSS Statistical Programme.*

This section aims to examine what the platforms most widely used by the sample subjects are. In *Table 3* it can be seen that 13 subjects have completed courses through various university platforms Moodle (6.5%), 3.5% in Coursera (7 subjects), 1.5% in Udacity (3 subjects) and 1% Uned COMA (2 subjects), a lower percentage is evident for platforms such as MiríadaX and EDX (0.5% with only one subject having used it).

As for the interaction of subjects in forums during MOOC courses 7.5% (14 subjects) reported no interaction while 3.5% (6 subjects) read forum conversations without participating in the interaction themselves.

Regarding the participation of the subjects in our sample study there is a greater number who answer question 12, because the questionnaire indicates that those unfamiliar with the MOOC phenomenon must go directly to that question number.

As can be seen in this case we have established four categories: in the first 6.5% of those in the study sample state that they have participated in these courses but have not finished any. In the second category, there are 14 subjects who participated and completed one or more courses (7%), while 6.5% of subjects answered in the negative with regard to participation or interest in the courses, with the remaining 3.5% saying that they did not participate but were interested in enroling.

|  |  | MOOC knowledge (C) | | | |
|---|---|---|---|---|---|
|  |  | Frequency | Percentage | Valid percentage | Cumulative percentage |
| **Motivation** | Free | 15 | 7,5 | 37,5 | 37,5 |
|  | MOOC topic | 8 | 4,0 | 20,0 | 57,5 |
|  | Curiosity about MOOC | 10 | 5,0 | 25,0 | 82,5 |
|  | Important university | 3 | 1,5 | 7,5 | 90,0 |
|  | Easy access to materials | 2 | 1,0 | 5,0 | 95,0 |
|  | Improve career prospects | 1 | ,5 | 2,5 | 97,5 |
|  | Personal development | 1 | ,5 | 2,5 | 100,0 |
|  | *Total* | 40 | 20,0 | 100,0 |  |
| Missing values | *System* | 160 | 80,0 |  |  |
|  | Total | 200 | 100,0 |  |  |

Table 4. *Motivation to achieve MOOC courses.* Source*: Compiled by author from SPSS Statistical Programme.*

The last question on the questionnaire is based on research into the motivation for the subject to complete the courses. As can be seen in Table 4, there is a variety of motivations that cause subjects to enrol in MOOC courses. Among the 40 subjects who answered, 7.5% (15 subjects) took these courses because of their gratuity, 4% (8 subjects) responded that they were interested in the topic, 5% (10 people) were curious about MOOC, 1.5% (3 subjects) were influenced by the college where the course was offered, 1% (2 subjects) suggested that easy access to materials was a factor while only 0.5% (2 subjects) answered that taking these courses was done to improve employment prospects and a further 0.5% mentioned personal development.

A survey with the variables sex and MOOC knowledge on the one hand has been conducted, and with the variables current state (employee, part-time or full-time worker) and MOOC knowledge on the other. To this end, a test has been carried out to analyze the contingency coefficient and *Chi square*, so that the significance of the relationship between the two variables can be checked. The results showed no significant difference.

It can be specified that, based on results obtained in the contingency tables, more women than men are aware of the MOOC phenomenon, because women were the majority in the sample study. Despite these results however, and in accordance with Pearson´s *Chi-square* test: equality of conditions are accepted and no relevant differences between men and women are noted.

Also, of the 10 who were aware of the phenomenon for over a year, 9 are students, while only one is working part time. Of of the 21 who became familiar with it during the last year, 17 are students and 4 are working. 10 students and 3 part-time workers have discovered MOOC in the last six months. Finally, the three people who became familiar with it one month ago are all students. Consulting Pearson's Asymptotic Significance *Chi-square* a value of 0.681 is noted enabling the acceptance of the equality of conditions but not the remaining significant differences between current state and MOOC knowledge variables.

### 5. Discussion and Conclusions

Once the research was completed, it could de said that all the objectives sought were achieved and the main hypothesis demonstrated, but also all results could be rejected save one. Starting from the main objective, which was to investigate the impact upon and the knowledge possessed by future teaching professionals in Spain with regard to the MOOC phenomenon, it can be asserted that knowledge about MOOC is insufficient and the impact upon sample subjects was minimal.

Regarding the proposed secondary objectives, the level of participation that future teachers have in courses different from MOOC has been assessed as well as the period of time elapsed since the acquisition of knowledge of the phenomenon. Likewise, initial awareness of the phenomenon and how that knowledge was obtained has been examined. Similarly, the problems found by subjects when following MOOC courses have also been analyzed, delving into the language selected by future education professionals in doing these courses. As well, attention has been given to which of the MOOC platforms are mostly used and the degree of interaction with other students or professionals throughout the completion of these courses noted. Similarly, the participation of students in these courses has been evaluated alongside the motivation that leads the subjects to participate in MOOC courses, as has been presented above. The information obtained for these objectives can be determined in the tables presented in this document.

Therefore, the conclusions will be described following the structure of the main hypothesis and those arising from it. As mentioned earlier, the main hypothesis is confirmed, i.e. future professionals of education in Spain do not have knowledge of the MOOC phenomenon. This hypothesis is accepted because of the fact that from a sample of 200 subjects the phenomenon was only known by 47, of whom 27 have followed some MOOC courses while the remaining 20 have yet to follow any or show any interest in doing so. As for the hypotheses derived from the main one it can be said that:

1. The subjects studying online courses different to MOOC are younger. This hypothesis however has been rejected, as 48 out of the total of 65 who have followed courses different to MOOC, are aged between 18 and 24, while the remaining 17 are between 24 and 35 years old. Despite these figures however, the Chi-square test indicates that these differences are not significant.
2. There is a higher percentage of students who are aware of the MOOC phenomenon and a lower percentage who are working part time or full time. Out of the total of 47 people who are familiar with MOOC, 39 are students and only 8 are working part time. As in the above hypothesis, Pearson´s Chi-square test shows that the equality of conditions is accepted, and no significant differences between current state and MOOC knowledge variables exist, so this hypothesis is rejected.
3. MiriadaX is the most widely used Internet portal for undertaking MOOC courses in Spain according to the SCOPEO report (2013). From the results it is notable that in our sample group only 0.5% have used MiríadaX, so this hypothesis will be rejected by ignoring what was collected in the research theoretical framework. According to SCOPEO (2013), MiriadaX is undoubtedly the most widely used portal in Spain, but the same is not valid for specific cases as observed in our study (training teachers).
4. The main motivation for the realization of these courses is gratuity. This hypothesis is accepted because out of the 40 subjects who answered the question, the majority stated that their undertaking of the courses was due to their gratuity rather than other options such as the improvement of job prospects or personal development.

Therefore, on the whole, it can be said that in recent years social changes have caused subjects to carry out a continuous learning curve in adapting continually to professional requirements and transformations (Brown & Adler, 2008; Blewitt & Cullingford, 2013; Schütze & Slowey, 2013; Head, Van Hoeck & Garson, 2015). Although MOOC is not the only teaching option that has emerged in the field of new technologies, it has opened up a new range of possibilities and benefits in the field of non-formal learning, allowing access to information anywhere and any time and in any field of expertise (Gómez-Galán, 2013; Billington, & Fronmueller, 2013; Olcott, 2013; Gómez-Galán & Pérez-Parras, 2014). In this sense, this new mass, has opened up scenarios which facilitate research and innovation, especially at universities, where teachers are required to face new educational challenges and work in collaborative virtual environments which are targeted at cooperation and the exchange of knowledge.

Probably MOOC is not the only teaching option for the future, but its philosophy should certainly be present in educational methodology. This phenomenon is expanding very rapidly around the world, so we believe that knowledge of it and its dissemination is necessary for universities. Its correct understanding and proper functioning will fulfil its potential and empower its strengths which will enrich and contribute to the teaching-learning process. This implies that knowledge of MOOC must be present in teacher-training studies although, as we have seen, this has yet to be realized. We have offered a new perspective on teacher training and MOOC, in relation to other international studies (i.e., Guo, Fang, Liu & Zhou, 2014; Li, Luk & Jong, 2014; or Tan, Goh & Sabastian, 2014).

Once the study was completed and the conclusions noted a number of limitations regarding this research have been highlighted. The first is that this is only a case study focusing on several universities and regions in Spain (Madrid, Extremadura, and Andalusia), so it would be appropriate to extend the subject sample to other Spanish regions and universities. However, according to all sociological statistics (CIS, 2014) the selected regions are representatives of the whole of Spain. It would also be of considerable importance to research not only the training of teachers but also to assess professionals who are already teaching. The same parameters should apply into research done into active teachers with regard to their knowledge of MOOC and the motivation which led them to take these courses. In order to achieve this it would be necessary to conduct a more comprehensive study, one which delves into the nature of MOOC courses in greater detail and analyzes more variables.

In doing this a deeper look into the issue should be aimed for, one that can determine its global importance in the teaching of professionals. Thus, not only will a better understanding of MOOC courses be achieved but also the using of new teaching methodologies that have been created from the ICT revolution will be encouraged and their ideas and knowledge shared.

In summing up it can be said that given the importance that MOOC courses in higher education have alongside the possibilities offered, the courses are poorly understood or not even known by future teaching professionals in Spain, where MOOC has practically no influence. It is therefore urgent that MOOC is studied and analyzed as part of present and future teaching syllabuses. Today, because of their importance in society, integration of ICT´s in all educational processes and their constant updating is required. It is necessary for all future education professionals.